\documentclass[a4paper,english]{rnti}

\usepackage[T1]{fontenc}
\usepackage[utf8]{inputenc}

\usepackage{url}

\usepackage{graphicx}

\titrecourt{The Collaborative Business Intelligence Ontology (CBIOnt) }

\nomcourt{M. Fahad et al.}

\titre{ The Collaborative Business Intelligence Ontology (CBIOnt)  }

\auteur{
        Muhammad Fahad\affil{1},
        Jérôme Darmont\affil{1},
        Cecile Favre\affil{1}
        }

\affiliation{
    \affil{1}Univ Lyon, Univ Lyon 2, UR ERIC, France\\
          firstname.lastname@univ-lyon2.fr,\\
 }

\resume{%
In the current era, many disciplines are seen devoted towards ontology development for their domains with the intention of creating, disseminating and managing resource descriptions of their domain knowledge into machine understandable and processable manner. Ontology construction is a difficult group activity that involves many people with the different expertise. Generally, domain experts are not familiar with the ontology implementation environments and implementation experts do not have all the domain knowledge. We have designed Collaborative Business Intelligence Ontology (CBIOnt) for BI4People project. In this paper, we present CBIOnt that is OWL~2~DL ontology for the description of collaborative session between different collaborators working together on the business intelligent platform. As the collaborative session be-tween various collaborators belongs to some collaborative form, phase and re-search aspect, therefore CBIOnt captures this knowledge along with the collaborative session content (comments, questions, answers, etc.) so that one can inference various types of information stored on ontologies when required. In addition, it stores the location and temporal-spatial information about the collaboration held between collaborators. We believe CBIOnt serves as a formal framework for dealing with the collaborative session taken place among collaborators on the semantic Web.
}

\summary{%
À l'heure actuelle, de nombreuses disciplines se consacrent au développement d'ontologies dans leurs domaines respectifs, dans le but de créer, diffuser et gérer des descriptions de ressources de leurs connaissances de domaine de manière compréhensible et traitable par des machines. La conception d'ontologies est une activité de groupe complexe, qui implique de nombreuses personnes aux compétences différentes. Généralement, les experts du domaine ne sont pas familiers avec les environnements d'implémentation d'ontologies et les experts en implémentation n'ont pas toutes les connaissances du domaine. C'est pourquoi, dans le cadre du projet BI4people, nous avons conçu une ontologie collaborative pour l'informatique décisionnelle (CBIOnt). Dans cet article, nous présentons CBIOnt,  qui est une ontologie OWL~2~DL pour la description de sessions collaboratives entre différents collaborateurs travaillant ensemble. 
Étant donné que la session collaborative entre ces collaborateurs appartient à une forme, une phase et un aspect de recherche collaboratifs, CBIOnt capture donc ces connaissances, ainsi que le contenu de la session collaborative (commentaires, questions, réponses, etc.), afin de pouvoir inférer divers types d'informations stockées sur des ontologies lorsque cela est nécessaire. De plus, CBIOnt stocke la localisation et les informations spatiotemporelles de la collaboration. Nous pensons ainsi que CBIOnt peut servir de cadre formel pour traiter une session collaborative qui a eu lieu entre des collaborateurs sur le Web sémantique.

}

\begin{document}

%
\section{Introduction}
The terms artificial intelligence (AI) are used to illustrate a diversity of technologies referring to the creation of intelligent software or hardware able to learn and solve problems. According to \citet{Thormundsson} from Statista Research Department, the global AI software market is predictable to develop approximately 54 percent year-on-year in 2020, reaching a forecast size of 22.6 billion U.S. dollars. These technologies include machine learning, computer vision, and natural language processing (NLP), among others. Nowadays ontology is regarded as a foundation stone in AI software. \citet{Guarino} defines: “ontology is generally regarded as a designed artifact consisting of a specific shared vocabulary used to describe entities in some domain of interest, as well as a set of assumptions about the intended meaning of the terms in the vocabulary”. 

In the current era, many disciplines are seen devoted towards ontology development for their domains with the intention of creating, disseminating and managing resource descriptions of their domain knowledge into machine understandable and processable manner. There are number of ontologies available for couple of domains. Some are very large having thousands of concepts and have been used in very important fields of science. Such as Gene Ontology \citep{Emily}, Disease Ontology \citep{Lynn}, Computer Assisted Brain Trauma Rehabilitation Ontology - CABROnto \citep{Zikos}, etc., are developed by many reputed consortiums which provide core biological knowledge representation for the modern biologists. Various small and general ontologies comprising few number of concepts are Music ontology \citep{RaimondM}, Conference ontology \citep{Nuzzolese}, Smart City ontology \citep{Qamar}, etc. Besides these, there are some general multi-usage ontologies which can be reused by other ontologies for their requirements. Example of these ontologies are Timeline Ontology \citep{RaimondT}, GeoNames ontology \citep{Maltese}, FOAF ontology \citep{Vakaj}, etc. 

In this paper, we present the Collaborative Business Intelligence (CBI) Ontology (CBIOnt) developed for the BI4people self-service BI platform \footnote{ https://eric.univ-lyon2.fr/bi4people/ } . CBIOnt is designed in OWL 2 DL ontology \citep{OWLspec} for the description of collaborative session between different collaborators working together on the business intelligence platform. It stores various types of knowledge; such as collaborative session belongs to which specific CBI form, collaborative decision phase and research aspect, location (date/time) of collaboration held and information about collaborators, etc. 

The rest of the paper is organized as follows. In the section 2, we introduce Collaborative Business Intelligence platform and present some basic scenario so that the reader of this paper should have in his mind why/where we need an ontology for the CBI in BI4People project. Section 3 elaborates the collaborative BI ontology; its main hierarchy, properties in CBI ontology, link with other online ontologies, etc. Finally section 4 concludes this paper and hints at future research.

\section{Background and Scenario for the CBI Ontology}

\subsection{Collaborative Business Intelligence Platform}

The goal of Collaborative Business Intelligence Platform is to enables data engineers and collaborators to make data accessible and supports many features such as information sharing, collaborative decision-making, project annotation management, etc., beyond the individual boundaries of enterprises and stakeholders. With the help of this platform, collaborators (people, companies, organizations) can make collaborative OLAP which facilitate users to easily and selectively extract and query data in order to analyze it from different points of views. Collaborators can sort their problems and address challenges by responding flexibly and intelligently to the dynamic and unpredictable situations. Figure 1 illustrates an example of a scenario based on the education data from the site of INSEE \footnote{ https://www.insee.fr/en/accueil } where different collaborators can create a pictorial chart for the cube measuring educated population (named Pop2to5ans) of three cities using the filter. We can see Measure (i.e., Educated Population), Dimensions (i.e., Geographical Label (LIBGEO), Region (REG)), Filter (LIBGEO equals Paris, Lyon, Marseille, Nice).

\begin{figure}[t]
\begin{center}
 \includegraphics[width=12cm]{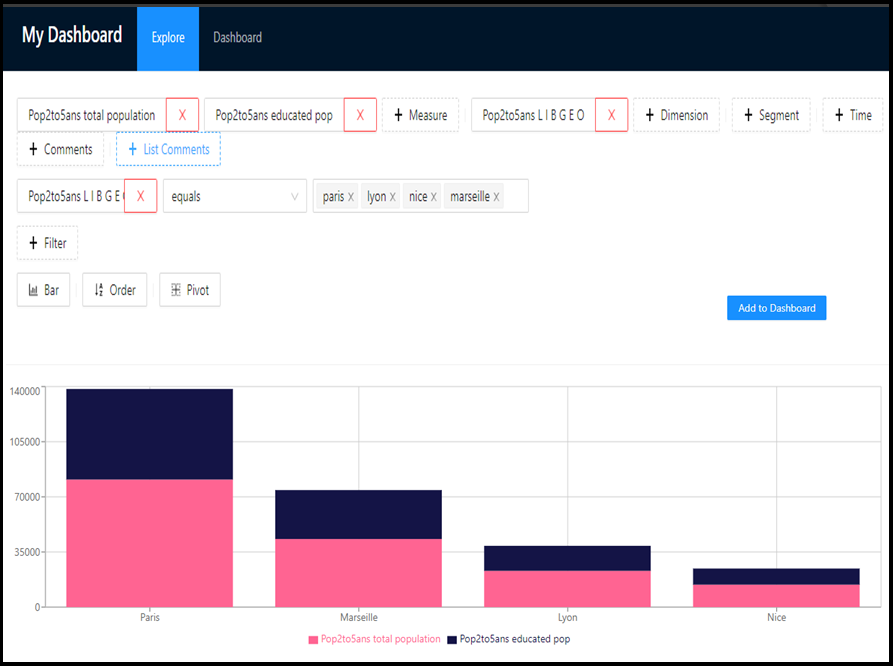}
 \caption{Bar Chart of population data in Collaborative BI Platform } \label{figure1}
\end{center}
\end{figure}

We implemented our collaborative OLAP by reusing CubeJS \footnote{ https://cube.dev/ } which is the head-less business intelligence platform. It has the necessary infrastructure to help data engineers and application developers to access data from modern data stores, organize it into consistent definitions, and deliver it to the application. CubeJS incorporates many features and implements efficient data modeling, access control, and performance optimization mechanisms so that an application can access consistent data via REST, SQL, and GraphQL APIs.

\subsection{Need for an ontology in the Collaborative BI platform }
In the current implementation of our Collaborative OLAP implementation, different collaborators can discuss and collaborate with each other. It supports adding annotations, comments, questions/answers, etc., on the dashboard during the collaborative session. One can edit already made dashboard and can add/update/remove questions, answers and comments along the pictorial illustration of data. Figure 2 illustrates an example of collaborative scenario where different collaborators add their comments on the dashboard. 
Initial implementation of BI platform stores these comments along the internal data and a feature is implemented where one can export collaborative session data including comments/questions/answers into a Jason file. But this technique of storage collaborative data is not efficient and effective for inference when there is a need of exploration of collaborative session data. Note that there is a question of storage and retrieval of this whole collaborative session and not just comments or question/answers noted in the session. We need to store all the information about the collaborative session, collaborators, content discussed, collaboration type, etc. Later this session data should be able to search and inference can be made for various purposes and information retrieval. Section 3 will elaborate what knowledge can be captured in the ontology and the concepts/properties of the designed ontology.

\begin{figure}[t]
\begin{center}
 \includegraphics[width=12cm]{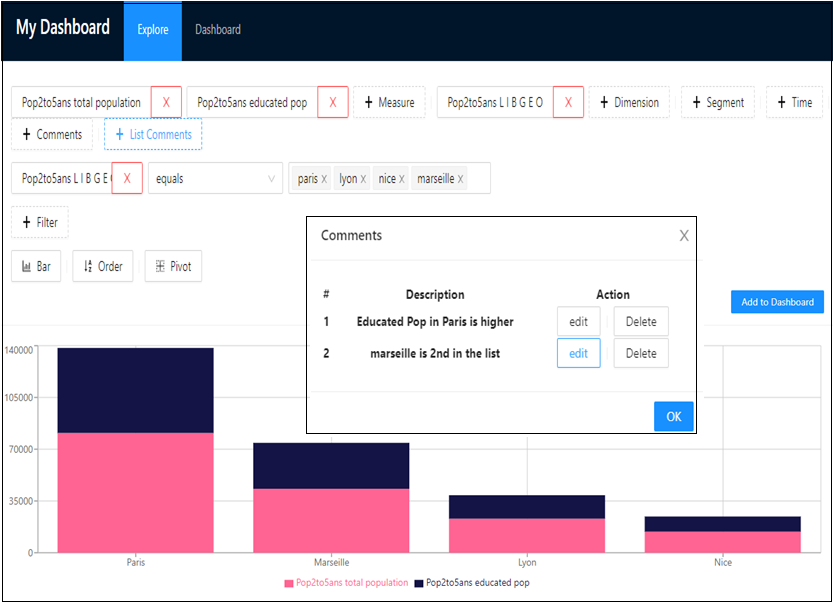}
 \caption{ Adding comments in the Collaborative BI platform } \label{figure2}
\end{center}
\end{figure}

\subsection{ Ontology Implementation in the Collaborative BI platform}
Engineering ontology for the collaborative BI is not easy and simple. First, ontology construction is a difficult group activity that involves many people with the different expertise. Generally, domain experts are not familiar with the ontology implementation environments (i.e., ontology notions, ontology development language, ontology editors and tool, etc.) and implementation experts do not have all the domain knowledge (concepts, relations, their links, etc.) to model. Second, the inherent problem with the real world for the complex systems (for example in our case collaborative session for the CBI platform) is the domain knowledge which is riddled with exceptions, complexity and difficulty during the extraction of appropriate necessary and sufficient conditions for the concepts from the domain facts of the real world data and complex situations.
For the Collaborative BI domain knowledge, we did a survey on collaborative business intelligence approaches, systems, and tools to come up with the necessary concepts to take into account in the engineering process of ontology for CBI.  In addition, we consider ontology web language OWL 2 DL as one of the best options available to engineer concepts and properties to capture CBI domain knowledge.

\section{Beginning of CBI Ontology}
Let us imagine a collaborative scenario where there is a collaborative session held between different collaborators at some location on some specific time. The collaborative session has many attributes and properties and has link to various objects of real world (see Figure 3). There are many questions that come across in our mind. Some of them are as follows.
\begin{itemize}
    \item Which location a collaborative session is held? 
    \item What data/time a collaborative session is held? 
    \item Which collaborators have participated in the session? Their affiliations?
    \item What is the main purpose of collaborative session? What are the outputs? 
    \item What is the main theme of collaboration? Which concepts it belongs to?
    \item What content is discussed during the session? Which topics are covered?
    \item Which topic or research aspect collaborative session is all about?
    \item What is the status of project? Which phase it belongs to? etc.
\end{itemize}

\begin{figure}[t]
\begin{center}
 \includegraphics[width=12cm]{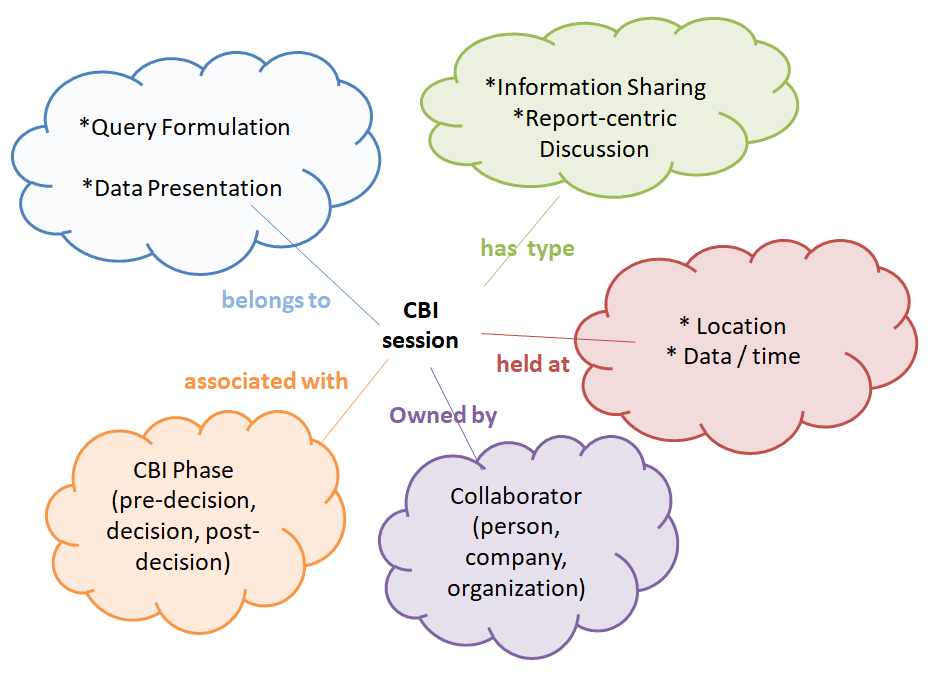}
 \caption{ Brainstorming CBI session main concepts } \label{figure3}
\end{center}
\end{figure}
    
To answer and capture information about these questions, we need to captures various types of information about the collaborative session such as collaborative research aspect, kind/forms of CBI topic, phase of project under CBI, etc. Figure 4 shows the main concepts of CBI session which form the main class hierarchy of CBI ontology. Following we discuss the main concepts and their hierarchies.

\begin{figure}[t]
\begin{center}
 \includegraphics[width=12cm]{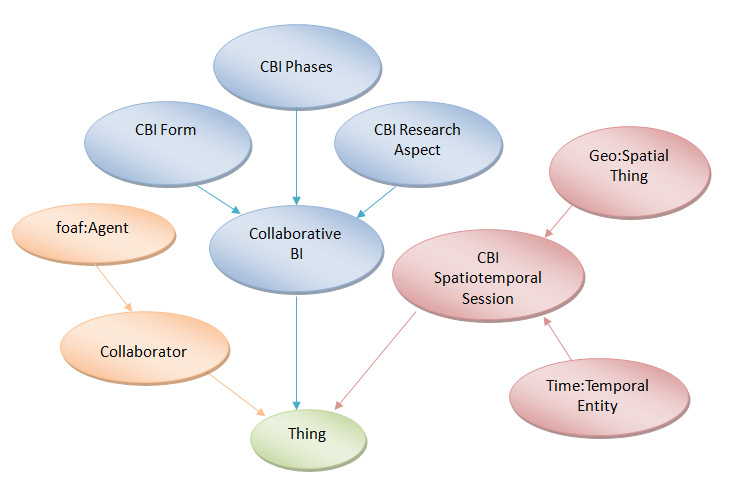}
 \caption{ The main hierarchy of CBI form concept. } \label{figure4}
\end{center}
\end{figure}

\paragraph{OWL Class : CBI\_Form.}
There are many forms of CBI. A very simple form of CBI can be a general discussion between different collaborators (i.e., people, companies, etc.), or it can be a report-centric discussions which can produce a re-port investigating a particular question. Collaborators can add annotations to specific items in a report to have more arguments between them. It can be analysis, reporting and visualizing customer behavior or enterprise trend, information sharing support and task coordination support, etc. Figure 5 illustrates the hierarchy of various forms of CBI under CBI\_Form concept. Therefore OWL Classes are generated for the following concepts {OWLClass:General\_Discussion, OWLClass:Annotation, OWLClass:Report\_ Centric\_Discussion, OWLClass:Visualizing\_Behavior, OWLClass:Trend\_Analysis, OWLClass:Task\_Coordination, OWLClass:Information\_Sharing} as subclasses under OWLClass: CBI\_Form.

\begin{figure}[t]
\begin{center}
 \includegraphics[width=12cm]{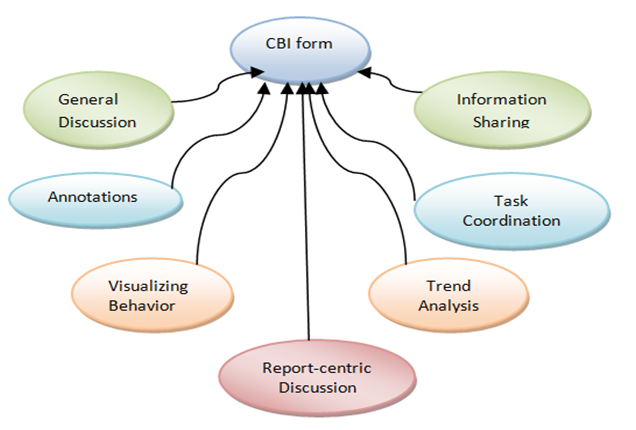}
 \caption{ CBI Form concept and its sub-class concepts } \label{figure5}
\end{center}
\end{figure}

\paragraph{OWL CLASS : CBI\_Research Aspect.} There are many research aspects of collaborative BI. \citet{Abello} listed five major sub-areas  of collaborative business intelligence such as query formulation, source discovery, and acquisition, integration and presentation of data. Collaborative query formulation where the user designs a query extended with a keyword-based search for the situational data based on the previous actions of other users in the same or similar cases. Another aspect can be collaborative source discovery or search of relevant data of high quality based on the tags provided by other users. When the desired data are not available online, the community can be asked for data; where others can upload desired data to be used by the whole community, known as collaborative data acquisition. After this phase, there can be a case where a user might want to add a location dimension to the cube. This can be accomplished by reusing existing parts available online or by collaboration of other users, also known as collaborative data integration. When the presentation is built on existing works within the so-called meta-morphing for BI analytical frontends is called the collaborative data presentation. Therefore these concepts are added as OWL Classes in the CBIOnt ontology under CBI\_Research\_Aspect concept as illustrated in Figure 6.

\begin{figure}[t]
\begin{center}
 \includegraphics[width=12cm]{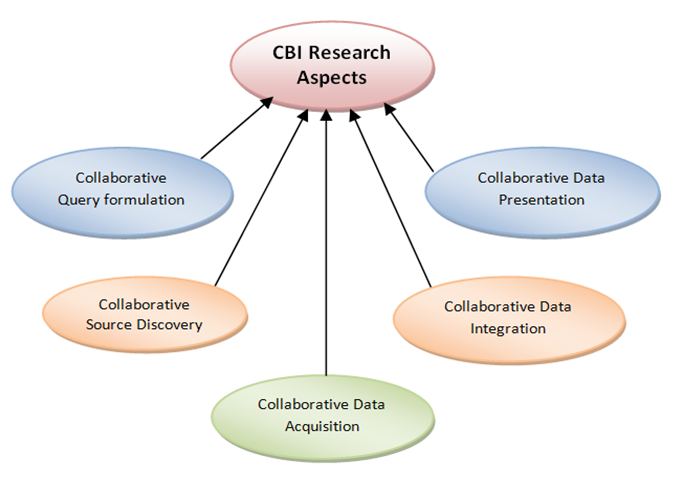}
 \caption{ CBI Research Aspect concept and its sub-class concepts } \label{figure6}
\end{center}
\end{figure}

\paragraph{OWL CLASS : CBI\_Phase.} There is a three phase process model of collaborative decision making process \citep{Adla}. Pre-decision phase restricts all participants to share understanding of the problem and the targeted objectives to achieve in the collaborative project. It assists to get a common description of the problem with respect to different viewpoints and also describes the limits and boundaries of the problem. There are four steps in the decision phase. Firstly, collaborators and participants generate their ideas and make a comparison of their ideas. Secondly, participants organize these ideas to promote the visibility and understanding of their ideas. Thirdly, participants compare these ideas in the different viewpoints for the common referential of the decision. Fourthly, collaborators identify and publish the agreements from participants and select one from all proposed solutions. Post-decision phase is very significant because it lets the monitoring of a decision. Monitoring a decision consists essentially to realize an action planning in order to implement the decision made. We have designed this ontology in web protégé which is also a collaborative ontology editor tool to engineer domain knowledge as ontologies. Figure 7 is taken from web protégé  and illustrates these phases as OWL Classes. 

\begin{figure}[t]
\begin{center}
 \includegraphics[width=12cm]{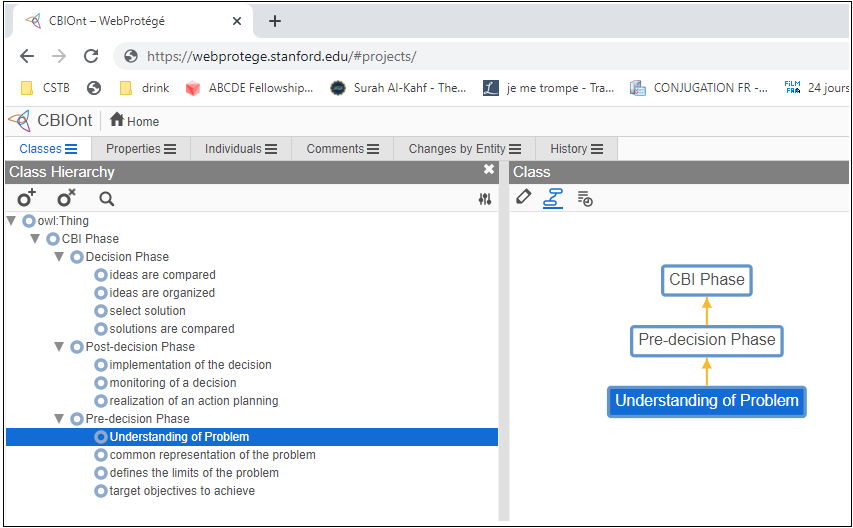}
 \caption{ CBI Phase concept and its sub-class concepts } \label{figure7}
\end{center}
\end{figure}

\paragraph{Data and Object Properties in CBIOnt Ontology}
There are three types of properties in OWL 2 ontology, i.e., Data Properties, Object Properties and Annotation Properties. According to OWL 2 web ontology language specification, Data properties connect individuals with literals and Object Properties connect pairs of individuals. Annotation properties can be used to provide an annotation for an ontology, axiom, or an IRI. In CBIOnt ontology, following are some of the important data and object properties:
\begin{itemize}
   
\item owned\_by: The object property owned\_by(CBI\_Session and Collaborators) represents the owner relationship between individuals of CBI\_Session and Collaborators (i.e., Person, Company, Organisation).
\item	associated\_with: The object property associated\_with(CBI\_Session and CBI\_Phase) represents that each collaborative session is associated with some CBI Phase (i.e., Pre-decision Phase, Decision Phase or Post-Decision Phase).
\item	belongs\_to: The object property belongs\_to(CBI\_Session and Research\_Aspect) represents that each collaborative session belongs to some research aspect, i.e., Collaborative Query Formulation, Collaborative Data Presentation, etc.
\item	has\_type: The object property has\_type(CBI\_Session and CBI\_Type) shows that each collaborative session belongs to some type of CBI, i.e., Information Sharing, Visualizing\_Behavior, General\_Discussion, etc. 
\item	hasRemark. CBI Session has very important Data properties also called attributes which captures the values of collaborative session. In OWL, one can use enumerations of literals when we need to specify the range of a datatype property. In CBIOnt ontology, hasRemark is an enumerations of literals that specify the range values {“Question”,“Answer”,“Comment”}.
\item	hasDescription. hasDescription is a data property that is used to associate text description of type String with the remarks generated in collaborative session. It captures the actual textual description of question/answer/comment asked during the collaborative session among different collaborators.
\end{itemize} 
Above are some of the important concepts and properties in CBI ontology. Later on, this ontology can operate as a foundation for more domain-specific knowledge representation and act as a backbone in building inference mechanisms in CBI system.

\section{Online Connected Ontologies and Concepts}
One of the ways to aid ontology engineering is to reuse existing ontologies to get a new ontology with the elaborative content and better quality. To build our ontology, we have reused some of the concepts from online ontologies. Following we discuss why and where we make links with the online ontologies and their concepts.

\paragraph{ The TimeLine ontology.} 
The Timeline ontology defines concept TimeLine which captures a coherent backbone for addressing temporal information \citep{RaimondT}. In addition, the TimeLine ontology uses two concepts (Interval and Instant) which are defined in OWL-Time \footnote{ https://www.w3.org/TR/owl-time/\#time:TemporalEntity}.  The time intervals are representing by concept Interval and instants are represented by Instant concept. Collaborative session must be held at some specific day and time. Therefore temporal information is vital to store and express when capturing knowledge for the collaborative session.  We build our CBIOnt on top of the timeline ontology to express and store temporal information. Therefore, we designed a concept CBI Temporal-Spatial Session which is a super-class of Time:TemporalEntity concept (i.e., class TemporalEntity extends Temporal-Spatial Session class).

\paragraph{ The GeoNames Ontology.} 
GeoNames \footnote{ https://www.geonames.org/ontology/documentation.html } is a well-known geospatial dataset providing data and metadata of around 7 million unique named places collected from several sources \citep{Maltese}. Collaborative session must be held at some location may be physical or virtual. Therefore, we designed a concept CBI Temporal-Spatial Session which is a super-class of Geo:SptialThing concept (i.e., class SptialThing extends Temporal-Spatial Session class ).

\paragraph{ The FOAF Ontology.} 
FOAF stands for Friend Of A Friend and FOAF ontology \footnote{ https://sparontologies.github.io/foaf/current/foaf.html }  is a machine readable ontology describing person, their activities and their relations to other people and objects \citep{Vakaj}. A FOAF agent can be a person, group, software or physical artifact, having subclasses {foaf:Group, foaf:Organization, foaf:Person}. Collaborative session must be among many collaborators which can be a person, company, organization, etc. Therefore, we designed a concept Collaborator which is a super-class of Foaf:Agent concept (i.e., class Agent extends Collaborator class).  

\section{Conclusion}
Ontologies have been developed and used widely across several disciplines and life sciences to build expert and intelligent AI systems. On the one hand, ontologies formally declare and conceptualize the domain knowledge to capture the real world objects, and on the other hand ontologies play a significant role in improving the accuracy, decision support and performance of entire systems. In this paper, we have presented our designed ontology for Collaborative Business Intelligence named CBIOnt. It has many concepts related to collaborative aspects so that it can store collaborative session data efficiently. It links several online ontologies and reuses their concepts. We believe that the semantic layer based on ontologies will play a major role in the development and advancement of CBI. It will help to improve the performance, effectiveness and precision of results as compared to the storage mechanisms in text file or relational table.
\section{Acknowledgment}
The research study depicted in this paper is funded by the French National Research Agency (ANR), project ANR-19-CE23-0005 BI4people.

\bibliographystyle{rnti}
\bibliography{biblio_exemple}

\providecommand\Fr{}
\providecommand\Eng{}
\providecommand\andname{and}
\providecommand\andnamec{and}

\begin{thebibliography}{}


\bibitem[{Abelló et~al.}(2013){Abelló, Darmont, Etcheverry, Golfarelli,
  Mazón, Naumann, Pedersen, Rizzi, Trujillo, Vassiliadis, \andnamec{}
  Vossen}]{Abello}
Abelló, A., J.~Darmont, L.~Etcheverry, M.~Golfarelli, J.-N. Mazón,
  F.~Naumann, T.~Pedersen, S.~Rizzi, J.~Trujillo, P.~Vassiliadis, \andname{}
  G.~Vossen (2013).
\newblock Fusion cubes: Towards self-service business intelligence.
\newblock {\em International Journal of Data Warehousing and Mining\/}~{\em 9},
  66--88.

\bibitem[{Adla et~al.}(2010){Adla, Nachet, \andnamec{} Ould-mahraz}]{Adla}
Adla, A., B.~Nachet, \andname{} A.~Ould-mahraz (2010).
\newblock Multi-agents model for web-based collaborative decision support
  systems.

\bibitem[{Bock}(2009){Bock}]{OWLspec}
Bock, Fokoue, H. H. H. R. S.~S. (2009).
\newblock Owl 2 web ontology language structural specification and
  functional-style syntax.
\newblock Technical report,
  https://www.w3.org/2007/OWL/draft/ED-owl2-syntax-20090531/.

\bibitem[{Emily et~al.}(2008){Emily, Barrell, Binns, Draghici, Camon, Hubank,
  Talmud, Apweiler, \andnamec{} Lovering}]{Emily}
Emily, Huntley, R., D.~Barrell, D.~Binns, S.~Draghici, E.~Camon, M.~Hubank,
  P.~Talmud, R.~Apweiler, \andname{} R.~Lovering (2008).
\newblock The gene ontology — providing a functional role in proteomic
  studies.
\newblock {\em PROTEOMICS\/}~{\em 8}.

\bibitem[{Guarino}(2000){Guarino}]{Guarino}
Guarino, N. (2000).
\newblock Formal ontology in information systems (fois).
\newblock {\em Proceedings of Formal Ontology and Information Systems,
  (FOIS'98): IOS Press\/}, 3--15.

\bibitem[{Lynn et~al.}(2011){Lynn, Arze, Nadendla, Chang, Mazaitis, Felix,
  Feng, \andnamec{} Kibbe}]{Lynn}
Lynn, S., C.~Arze, S.~Nadendla, Y.-W.~W. Chang, M.~Mazaitis, V.~Felix, G.~Feng,
  \andname{} W.~Kibbe (2011).
\newblock Disease ontology: A backbone for disease semantic integration.
\newblock {\em Nucleic acids research\/}~{\em 40}, D940--6.

\bibitem[{Maltese \andnamec{} Farazi}(2013){Maltese \andnamec{}
  Farazi}]{Maltese}
Maltese, V. \andname{} F.~Farazi (2013).
\newblock A semantic schema for geonames.
\newblock {\em In proceedings of INSPIRE Conference – June 25th 2013\/}.

\bibitem[{Nuzzolese et~al.}(2016){Nuzzolese, Gentile, Presutti, \andnamec{}
  Gangemi}]{Nuzzolese}
Nuzzolese, A., A.~Gentile, V.~Presutti, \andname{} A.~Gangemi (2016).
\newblock Semantic web conference ontology - a refactoring solution.
\newblock Volume 9989, pp.\  84--87.

\bibitem[{Qamar et~al.}(2019){Qamar, Bawany, Javed, \andnamec{} Amber}]{Qamar}
Qamar, T., N.~Bawany, S.~Javed, \andname{} S.~Amber (2019).
\newblock Smart city services ontology (scso): Semantic modeling of smart city
  applications.
\newblock pp.\  52--56.

\bibitem[{Raimond et~al.}(2007){Raimond, Abdallah, Sandler, \andnamec{}
  Giasson}]{RaimondM}
Raimond, Y., S.~Abdallah, M.~Sandler, \andname{} F.~Giasson (2007).
\newblock The music ontology.

\bibitem[{Raimond \andnamec{} S.}(2007){Raimond \andnamec{} S.}]{RaimondT}
Raimond, Y. \andname{} A.~S. (2007).
\newblock The timeline ontology - owl-dl ontology.
\newblock Technical report,
  http://motools.sourceforge.net/timeline/timeline.html.

\bibitem[{Thormundsson}(2022){Thormundsson}]{Thormundsson}
Thormundsson, B. (2022).
\newblock Artificial intelligence software market growth forecast worldwide
  2019-2025.
\newblock Technical report,
  https://www.statista.com/statistics/607960/worldwide-artificial-intelligence-market-growth/.

\bibitem[{Vakaj \andnamec{} Martiri}(2011){Vakaj \andnamec{} Martiri}]{Vakaj}
Vakaj, E. \andname{} E.~Martiri (2011).
\newblock Foaf-academic ontology: A vocabulary for the academic community.
\newblock pp.\  440--445.

\bibitem[{Zikos et~al.}(2013){Zikos, Galatas, Metsis, \andnamec{}
  Makedon}]{Zikos}
Zikos, D., G.~Galatas, V.~Metsis, \andname{} F.~Makedon (2013).
\newblock A web ontology for brain trauma patient computer-assisted
  rehabilitation.
\newblock {\em Studies in health technology and informatics\/}~{\em 190},
  100--102.

\end{thebibliography}

\end{document}